\documentclass{aa}
\usepackage{graphicx}
\usepackage[figuresright]{rotating}
\usepackage{aalongtable,lscape}

\begin{document}

   \title{The Circumnuclear Star-forming Activities along the Hubble Sequence}

   \author{Lei Shi, Qiusheng Gu \and Zhixin Peng}

   \institute{Department of Astronomy, Nanjing University,
              Nanjing 210093, China\\
              \email{shileiastro@yahoo.com; qsgu@nju.edu.cn}
             }

  \date{Received ----; accepted ----}

   \abstract{In order to study the circumnuclear star-forming
   activity along the Hubble sequence, we cross-correlate the
   Sloan Digital Sky Survey Data Release 2 (SDSS DR2) with the
   Third Reference Catalog of Bright Galaxies (RC3) to derive a
   large sample of 1015 galaxies with both morphological and
   spectral information. Among these, 385 sources are classified
   as star-forming galaxies and the SDSS fibre covered the
   circumnuclear regions (0.2 $-$ 2.0 kpc). By using the spectral
   synthesis method to remove the contribution from the underlying
   old stellar population, we measure the emission lines fluxes
   accurately which are then used to estimate the star formation
   rates(SFRs). Our main findings are that: (1) Early-type spirals
   show much larger H$\alpha$ luminosities and hence higher SFRs,
   they also suffer more extinctions than late-type ones. The
   equivalent widths (EWs) of H$\alpha$ emission lines show the similar
   trend, however, the very late types (Sdm $\sim$ Irr) do have
   large fractions of high EWs. (2) We confirm that D$_n(4000)$
   has strong correlation with the strengthes of metallic
   absorption lines (such as CN band, G band and Mg Ib). Both
   these lines and the Balmer absorption lines show interesting
   variations between Sbc to Sd type galaxies. (3) The bar
   structure tightly relates with the enhanced star formation
   activity, this effect is even more significant in the
   early-type spirals. But we should note that the bar structure
   is not a necessary or sufficient condition for galaxies to
   harbor circumnuclear star formations.

   \keywords{ galaxies: general -
              galaxies: stellar content -
              galaxies: statistics
               }
   }

   \titlerunning{Star-forming activity along the Hubble sequence}

   \authorrunning{Shi, et al.}

   \maketitle
%

\section{Introduction}

    The star formation history is one of the most important
    problems for understanding the evolutions of the galaxies and
    also the universe. Enormous progress have been made during the
    last two decades. More precise diagnostics of global star
    formation rates (SFRs) have been obtained in a broad range of
    techniques: e.g., integrated optical emission line fluxes
    (Kennicutt 1983; Kennicutt et al. 1994; Madau et al. 1998),
    near-ultraviolet continuum fluxes (Donas \& Deharveng 1984;
    Deharveng et al. 1994; Madau et al. 1998; Bell \& Kennicutt
    2001), infrared continuum fluxes (Harper \& Low 1973; Rieke \&
    Lebofsky 1978, Telesco \& Harper 1980; Kennicutt et al. 1987),
    and radio emissions (Condon 1992; Cram et al. 1998). Kennicutt
    (1998) has presented an excellent review on star formation
    along the Hubble sequence.

    A galaxy's morphology carries an important information for the
    study of star formation history in galaxies (e.g., Kennicutt
    1998). Hubble sequence (Hubble 1926), as one of the most
    widely used classification of galaxies, is closely connected
    with the global star formation activities of galaxies. The
    basic view of global SFRs along the Hubble sequence shows that
    the equivalent widths (EWs) of H$\alpha$ emission lines
    increase from virtually zero in E/S0 galaxies to about a
    couple of hundreds \AA\ in late-type spiral and irregular
    galaxies (Kennicutt \& Kent 1983; Kennicutt 1998). However,
    recent studies of H$\alpha$ imaging indicate that a
    significant fraction of the early-type spirals shows
    remarkable star formation activities ( Young et al. 1996;
    James et al. 2004; Hameed \& Devereux 2005).

    Many galaxies are shown to harbor luminous star forming
    activities at the circumnuclear region with properties different
    from those of extended disks (Morgan 1958, S\'ersic \& Pastoriza
    1967). Comprehensive surveys of the star formation properties of
    galactic nuclei have been carried out both in the optical band
    (Stauffer 1982, Keel 1983, Kennicutt et al. 1989, Ho et al. 1997a,
    b) and mid-IR photometry (Rieke \& Lebofsky 1978, Scoville et al. 1983,
    Devereux et al. 1987, Giuricin et al. 1994). Ho et al. (1997a) found
    that the fraction of nuclear emission spectra with HII region-like
    line ratios increased from virtually zero in elliptical galaxies,
    8\% in S0 galaxies to 80\% in Sc-Im galaxies. However, these
    could be contaminated by LINER or Seyfert nuclei.

    In most galaxies, the nuclear SFRs are quite modest, while the
    highest SFRs are mainly seen in the IR observations because the
    very luminous starbursts are mostly associated with dense molecular
    gas and suffered substantial extinctions in the optical band
    The details of these IR luminous circumnuclear properties could
    be seen in Veilleux et al. (1995),
    Lutz et al. (1996), and also in Kennicutt (1998).

    In this paper, we cross-correlate the largest spectral
    database available from the Sloan Digital Sky Survey (SDSS)
    and the Third Reference Catalog of Bright Galaxies (RC3) to
    derive a large sample of galaxies with both morphological and
    spectral information, we thus are able to statistically study the
    star forming activities along the Hubble sequence. This paper
    is organized as follows: we will describe the sample selection
    and also give a brief introduction of the stellar synthesis model
    in Section 2; the statistical results are presented in Section 3;
    in Section 4, we will discuss bar influence on the star formation
    activities; finally, we present our main conclusions in Section 5.


\section{Sample}

    The Sloan Digital Sky Survey (SDSS, York et al. 2000;
    Stoughton et al. 2002; Abazajian et al. 2003) is one of the
    most ambitious surveys, which aims to obtain nearly $10^6$
    galaxies and $10^5$ quasars when completed. This provides us a
    great opportunity to explore the relations between the
    star-forming properties and their morphologies. We thus
    cross-correlate the SDSS galaxies (Data Release 2, Abazajian
    et al. 2004) with the RC3 catalog using the position-matching
    accuracy of 1 arc minute for most galaxies and 3 arcminutes
    for those without enough accurate positions in RC3 (see the
    introduction of RC3). We derived 983 galaxies with an
    average of positional differences of 0.0749 $\pm$ 0.052 arc
    min, 11 galaxies with 0.7246 $\pm$ 0.1716 arc min, and 33
    galaxies with 2.055 $\pm$ 0.9156 arc min. For those 44
    galaxies with large positional differences, we checked one
    by one by using Sloan Digital Sky Survey/SkyServer
    \footnote{http://cas.sdss.org/astro/en/tools/search/}.
    Finally, we excluded 12 galaxies for confusion and derived a
    sample of 1015 galaxies, 39 of which were defined as AGNs in
    V\'eron-Cetty \& Veron (2003) and excluded for further
    analysis. We derived consistent results by comparing with NYU
    Value-Added Galaxy Catalog (Blanton et al. 2005)
    \footnote{http://wassup.physics.nyu.edu/vagc/}, where they
    have presented SDSS galaxies catalogue with RC3 information.

    Since the SDSS spectrophotometry used the size-fixed fibre (3
    arcseconds), according to the the physical covering size of
    the SDSS fibre, we divided the whole sample into three
    sub-samples which mainly depends on the SDSS fibre covering
    region size: (1) 33 nuclear sub-sample, the fiber covering the
    region of less than 200 pc; (2) 824 circumnuclear sub-sample,
    where the fiber covers region of 0.2 kpc $\sim$ 2 kpc; (3) 119
    disk sub-sample, covering the region larger than 2 kpc.
    Just like most, if not all, our sample will suffer from the
    Malmquist bias. In the following we will only concentrate on
    the circumnuclear sub-sample, with the relatively narrow
    redshift range of [0.00344,0.0344], and we will also use the
    distance-independent parameters (such as EWs and H$\alpha$
    luminosity normalized by the fibre covering size) as
    indicators of SFRs, both of which will reduce the Malmquist
    effect for this study.


    Though RC3 (de Vaucouleurs et al. 1991) has been wildly used
    in astrophysical researches, many excellent works have been
    carried out both in photometric and spectroscopic observations
    ever since then, which undoubtedly revised some galaxies'
    morphological classifications (such as de Souza, Gadotti \&
    dos Anjos 2004). As the NASA Extragalactic Database (NED)
    morphological types are the most up-to-date types in
    existence, we will use the NED morphologies for this work.



    The results of cross-correlation between RC3 and SDSS DR2 of
    all the 976 galaxies are shown in Table \ref{Table1}, which
    contains the galaxy name (Column 1); R.A. and Dec. (2000,
    Column 2-3); redshift (Column 4); stellar and nebular extinctions
    (Column 5-6); luminosity of far-IR, in unit of log erg s$^{-1}$
    (Column 7); morphology from NED (Column 8).

    \begin{table}[htb]
    \caption[Morphology frequency]{The frequencies with different
        morphological types for the circumnuclear sample}
    \label{tb1}
    \begin{center}
    \begin{tabular}{ccccccccc}
      \hline
        Galaxy Type   & E  & L  & S  & Irr & Uncertain & Total \\
      \hline
        Number & 43 & 127 & 536 & 18 & 100 & 824 \\
        Percentage & 5.2 & 15.4 & 65.0 & 2.2 & 12.1 & 100 \\
      \hline
      \end{tabular}
    \end{center}
    \begin{list}{}{}
    \item[] All galaxies were obtained from the cross-certification
        of RC3 and SDSS DR2 with the redshifts ranging from 0.0034 to
        0.0344. The types are taken from NED calculated by using the
        definition of de Vaucouleurs et al. (1991).
    \end{list}
    \end{table}

    Here we have combined all the elliptical galaxies into index
    T = $-$2, and included peculiar ones to Irregular galaxies (T
    = 10) for simply assuming that they may all involved in
    various merger processes. The index T = 13 are set for the
    uncertain or unusual type galaxies and will be excluded from
    the statistical studies. Table \ref{tb1} gives the number of
    each main branches of Hubble sequences for the circumnuclear
    sample, which contains 43 ellipticals, 127 lenticulars, 536
    spirals and 18 irregulars. Just like most, if not all, galaxy
    samples, the current circumnuclear sample of galaxies can not
    be considered complete. We hope that the much larger number
    of galaxies could reduce the problem of incompleteness.

    In most normal galaxies, stellar light dominates the continuum
    and absorption lines at optical band. These absorption
    features can certainly contaminate the nebular emissions which
    will be useful for understanding the physical properties of
    galaxies. Moreover, the stellar spectrum alone may provide
    direct measurements on the stellar population, velocity
    dispersion and star formation history (SFH) etc. Thus how to
    properly decompose the stellar component from the integrated
    spectrum becomes the first step to interpret the nuclear
    properties.

    We use the same stellar population synthesis code provided by
    Roberto Cid Fernandes, called {\it starlight V2.0},  for
    spectra synthesis, which fits an observed spectrum $O_\lambda$
    with a linear combination of $N_\star$ simple theoretical
    stellar populations (SSP) computed with evolutionary synthesis
    models of Bruzual \& Charlot (2003, BC03). The methodology of
    the code is described in detail by Cid Fernandes et al.
    (2004b). In this work we adopt a base with $N_\star = 45$
    SSPs, with 3 metallicities: $Z = 0.2$, 1 and 2.5 $Z_\odot$,
    each of which has 15 ages ranged as 0.001, 0.003, 0.005, 0.01,
    0.025, 0.04, 0.10, 0.29, 0.64, 0.90, 1.4, 2.5, 5.0, 11 and 13
    Gyrs.  The match between model and observed spectrum is
    evaluated by the ruler of $\chi^2 = \sum_\lambda \left[
    \left(O_\lambda - M_\lambda \right) w_\lambda \right]^2$,
    where $w_\lambda^{-1}$ is the error in $O_\lambda$. The
    search for the best solution (also a minimum $\chi^2$ ) is
    carried out by using a simulated annealing plus Metropolis
    scheme, consisting of a series of 6 likelihood-guided
    Metropolis explorations through the parameter space (see Cid
    Fernandes et al. 2001 for a detailed discussion of the
    Metropolis method applied to the population synthesis
    problem), which searches for the minimum $\chi^2$. Spectral
    regions around emission lines, bad pixles or sky residuals are
    masked out by setting $w_\lambda = 0$.

    The spectra were first rebinned to 1 \AA\ bins and redshift
    corrected to the rest-frame by the standard IRAF \footnote{IRAF is
    distributed by the National Optical Astronomy Observatory, which
    is operated by the Association of Universities for Research in
    Astronomy, Inc., under cooperative agreement with the National
    Science Foundation} procedures. Then we corrected the spectra for
    Galactic Extinction using the reddening law of Cardelli et al
    (1989) and the A$_{B}$ values listed in NED (Schlegel, Finkbeiner
    \& Davis 1998) before starting the synthesis. We normalize the SSP
    basses at $\lambda_{0}$ = 4020 \AA, while for the observed spectra
    we use the median value between 4010 and 4060 \AA. Every spectrum
    has been synthesized from 3650 to 8000 \AA.

\begin{figure}
    \centering
    \includegraphics[width=8cm]{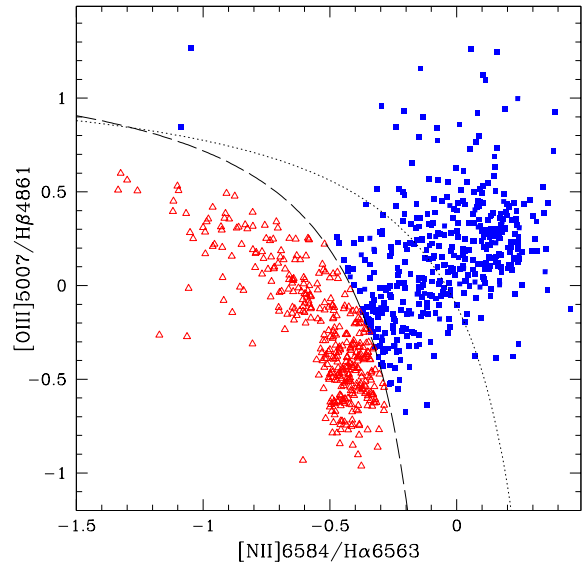}
    \caption[BPT Diagram]{The distribution of the galaxies in our
        circumnuclear sample in the BPT emission line ratio diagram.
        The dotted line is from Kewley et al. (2001), and the long-dashed
        line is from Kauffmann et al. (2003), which will be used in
        this paper as the criteria to distinguish star forming galaxies
        and the AGNs or the composite ones. The filled rectangles show
        the AGNs or the composite ones; while the open triangles
        represent the star forming galaxies. \label{fig2.3}}
    \end{figure}

\begin{figure}
    \centering
    \includegraphics[width=8cm]{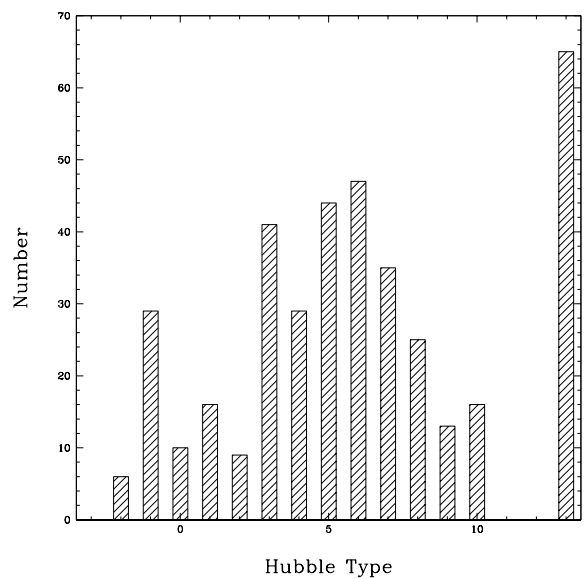}
    \caption[Sample division]{The morphological distribution for
    the 385 star-forming galaxies. We set the T-index of those galaxies with
    uncertain types as T = 13. \label{fig2.2}}
    \end{figure}

    Since in this paper we will only concentrate on the star
    forming galaxies, we use the criteria given by Kauffmann et
    al. (2003) to distinguish the potential AGNs or the composite
    ones from star-forming galaxies. Figure \ref{fig2.3} is the
    diagram of Baldwin, Phillips, \& Terlevich (1981, BPT)
    for our sample. After adopting the line of
    Kauffmann et al. (2003), we derived 385
    star forming galaxies and 438 AGNs or the composite ones.
    There is still one galaxy not presented in this diagram,
    because it shows no emission line features both in its
    original spectrum and the continuum subtracted one.
    Figure \ref{fig2.2} shows the morphological distribution for
    these star-forming galaxies.

%

%


\section{Statistic Results}

\subsection{The Emission Line Features}
    We have measured the emission line features from the "pure emission"
    spectra, which are derived by subtracting synthetic spectrum from
    the observed one. The measurement is automatical and the windows for
    emission line and continuum are listed in Table \ref{tb2}.

    \begin{center}
    \begin{table}[htb]
    \caption[Definitions of emission line windows]{Definitions of
        emission line windows}
    \label{tb2}
    \begin{tabular}{ccl}
    \hline
        Emission Lines & Windows   & Continuum Windows\\
    \hline
        H$\delta$ 4101 & 4090-4114 &    4055-4065 4135-4145\\
        H$\gamma$ 4340 & 4328-4352 &    4250-4260 4400-4410\\
        HeII 4686      & 4672-4700 &    4655-4665 4720-4730\\
        H$\beta$ 4861  & 4851-4871 &    4825-4835 4885-4895\\
        $[OIII]$ 4959  & 4949-4969 &    4915-4925 5035-5045\\
        $[OIII]$ 5007  & 4997-5017 &    4915-4925 5035-5045\\
        H$\alpha$ 6563 & 6556-6573 &    6520-6530 6610-6620\\
    \hline
    \end{tabular}

    \noindent All wavelengths are in units of \AA.
    \end{table}
    \end{center}
%
    \begin{figure}
    \centering
    \includegraphics[width=8cm]{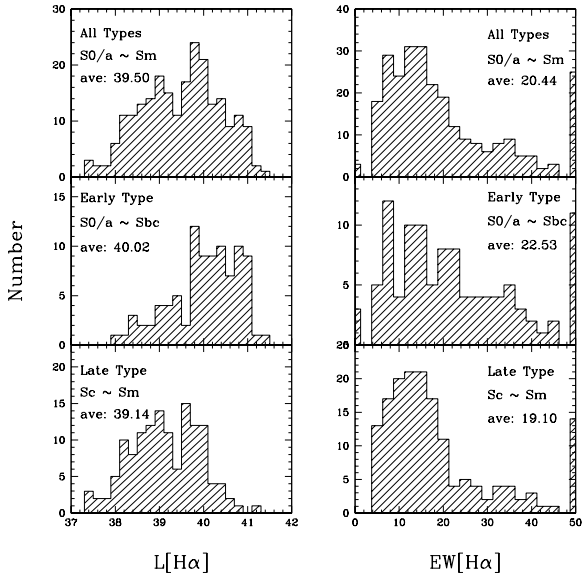}
    \caption[Comparison of L(H$\alpha$) and EW(H$\alpha$) between early and late type
        spirals]{The histogram of extinction-corrected L(H$\alpha$) and EW(H$\alpha$).
        From the top to bottom are the distributions of H$\alpha$ luminosities
        and equivalent widths for all, early-type and late-type galaxies.
        The luminosities have all been corrected for Balmer extinction.
        \label{fig4.2.1}}
    \end{figure}
    The H$\alpha$ luminosities of the sample span a large ragne, from
    about 10$^{37}$ to 10$^{42}$ erg s$^{-1}$ (Figure \ref{fig4.2.1}). The
    early-type (S0 $-$ Sbc) spirals extend much larger toward the high end
    of L(H$\alpha$) than the late-type (Sc $-$ Im) ones \footnote{From
    here to the end of this paper,
    the early-type refers to S0/a$-$Sbc and the late-type refers to
    Sc$-$Sm as the same in Ho et al. (1997a, b).}, which is
    consistent with Ho et al. (1997a). The median value of
    log(L[$H\alpha$]) is 40.02 and 39.14 for the early-type and
    late-type galaxies respectively. We should note that this trend is
    truly exist and slightly exaggerated by the effect of nebular
    extinctions. Table \ref{Table2} shows the flux densities of
    H$\beta$ 4861 (Column 2), [OIII] 5007 (Column 3), [NII] 6548,6584
    (Column 4-5), H$\alpha$ 6563 (Column 6), and [SII] 6717,6731 (Column 7-8)
    for our sample, all in units of $10^{-17} erg\ s^{-1}cm^{-2}$.
    Corresponding equivalent widths are shown in Table
    \ref{Table3} for all the sample galaxies, in units of \AA.

\subsection{Extinction}
    The dust extinction can have considerable
    impact on the thermal balance of the interstellar medium and on
    the formation of H$_2$ molecules, which have important consequences
    on the efficiency of star formation (Omukai 2000, Hirashita \&
    Ferrar 2002). The nebular extinction is derived from the observed
    Balmer decrement (H$\alpha$/H$\beta$, see Torres-Peimbert, Peimbert
    \& Fierro 1989). Assuming Case B
    recombination and a standard reddening law (Cardelli et al. 1989),
    we obtained the V-band extinction:
    \begin{equation}
        A_{V}^{nebular} = 6.31 \times\ \log\
        (\frac{F_{H\alpha}/F_{H\beta}}{I_{H\alpha}/I_{H\beta}}),
    \end{equation}

    \noindent where $F_{H\alpha}/F_{H\beta}$ and $I_{H\alpha}/I_{H\beta}$ are
    the observed and intrinsic flux ratios of H$\alpha$ and H$\beta$
    respectively. Here we use the intrinsic ratio of
    $I_{H\alpha}/I_{H\beta}$ to be 2.86 (Osterbrock 1989).
    Galaxies with either one of the two lines insufficient in fluxes
    (less than 3 $\sigma$) were excluded here. We also derived some
    negative values of A$_{V}$. Since these were unexpected for the
    common model of extinction and may actually be overestimated
    (see Cid Fernandes et al. 2005 for the details), we set them
    as zero in studying the relation between the extinctions and
    Hubble types, and also in the previous study of H$\alpha$
    luminosities.

    It is believed that emission lines suffer more extinctions
    than stellar light (Calzetti, Kinney \& Storchi-Bergmann 1994;
    Gordon et al. 1997; Mas-Hesse \& Kunth 1999). More recently,
    Cid Fernandes et al. (2005) have applied the synthesis model on
    a large number of SDSS galaxies and get a linear bisector
    fitting $ A_V^{\rm Balmer}  = 0.24 + 1.81 A_V $. Since we use
    the same code as Cid Fernandes et al. (2005), it is not
    surprised that we get similar results.

    \begin{figure}
    \centering
    \includegraphics[width=8cm]{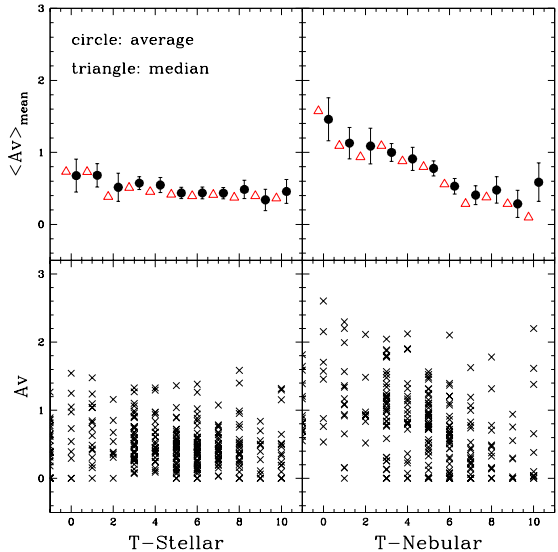}
    \caption[The distribution of extinction along Hubble sequence]{The
        distribution of extinction along the Hubble sequence. The left two
        panels are the distributions of stellar extinctions (bottom panel)
        and their median (open triangles) and average values
        (filled circles). The error bar represents the deviation in the
        mean. The right plots are for the nebular extinctions.
    \label{fig4.1.2}}
    \end{figure}

    Figure \ref{fig4.1.2} shows the distribution of both the stellar and
    nebular extinctions ($A_V$) along the Hubble sequence. We find
    that the nebular extinction decreases from early-type spirals to
    late-type spirals, which
    is consistent with Ho et al. (1997a) and Stasi\'nska et al. (2004).
    while the trend of the stellar extinction is almost flat with slight
    variations among morphology types. This is accessible, since the
    stellar extinction represents the comparably old stars which
    are believed to have exposed from the surrounding molecular
    clouds.

\subsection{Star Formation Rates}

\begin{figure}
    \centering
    \includegraphics[width=8cm]{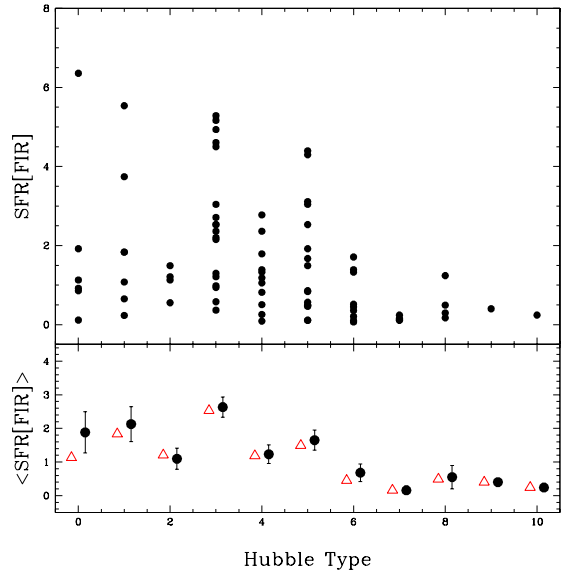}
    \caption[The circumnuclear SFRs along Hubble sequence]{The
        circumnuclear SFRs of H$\alpha$ along the Hubble sequence. Most
        galaxies show moderate SFRs, while large dispersions are seen
        in early-type galaxies. The median SFR values show a clear trend
        decreasing from early-type to late-type spirals. Symbols are the
        same meaning as Figure \ref{fig4.1.2} \label{fig4.3.1}}
    \end{figure}

\begin{figure}
    \centering
    \includegraphics[width=8cm]{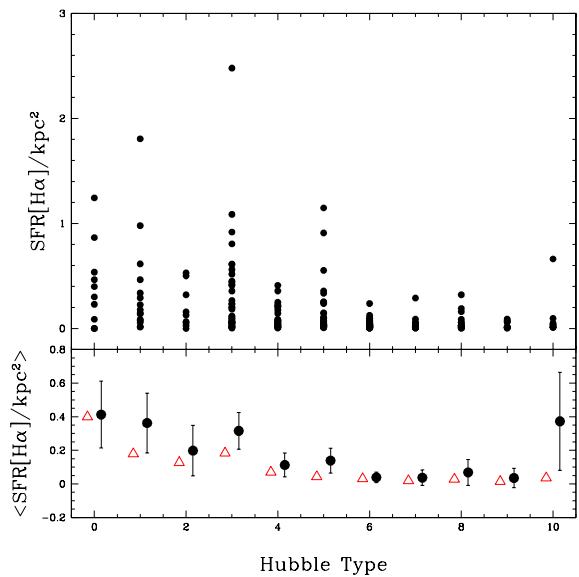}
    \caption[The distribution of SFRs normalized by aperture size]{The
        distribution of SFRs normalized by the aperture areas. The trend
        is much weaker, however, the early-type galaxies do show higher
        median values of SFR(H$\alpha$). \label{fig4.3.5}}
    \end{figure}

\begin{figure}
    \centering
    \includegraphics[width=8cm]{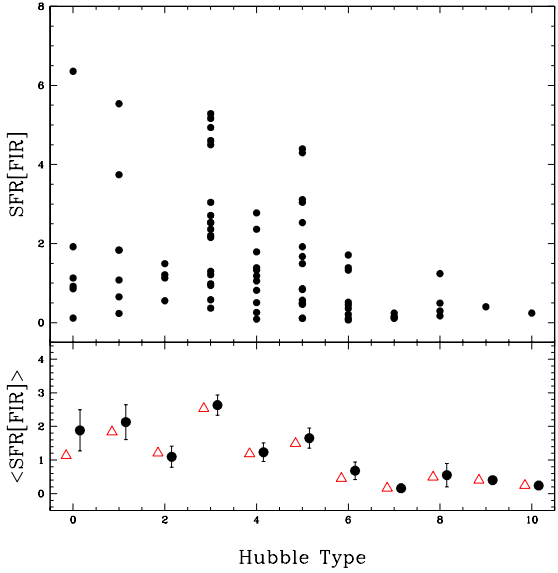}
    \caption[Far-Infrared star formation rates along Hubble
        sequence]{The distribution of FIR SFRs among Hubble types.
        The early-type spirals show comparably larger values,
        however, this may due to the lack of data in late-type
        spirals. Symbols are the same meaning as Figure
         \ref{fig4.1.2}\label{fig4.3.2}}
    \end{figure}

    We have calculated the star formation rate from the
    extinction-corrected H$\alpha$ luminosity using Kennicutt's (1998)
    conversion,
    \begin{equation}
        {\rm SFR}~(M_\odot~yr^{-1}) = 7.9 \times 10^{-42}~L(H\alpha)~({\rm
        ergs~s^{-1}}),
    \end{equation}
    Here we get an average SFRs of 0.14 $M_{\odot}$ $yr^{-1}$ with the
    median value of 0.04 $M_{\odot}$ $yr^{-1}$ for those with A$_{V}$
    or no sufficient flues in both H$\alpha$ and H$\beta$ lines which
    would be set as zero in SFRs of our 385 galaxies. These values are consistent with Ho
    et al. (1997a). Figure \ref{fig4.3.1} (top) shows the distribution
    of star formation rates as a function of Hubble T-types. We find
    that the spiral galaxies of type Sb show the highest star formation
    rate, which is slightly different from the global properties of the
    H$\alpha$ survey (James et al 2004, peaking at Sbc and Sc). There
    exists a clear trend along the Hubble types with the values of
    SFR(H$\alpha$) decreasing from early-type spirals to the late-type
    ones (Figure \ref{fig4.3.1}, bottom). This is consistent with Stauffer
    (1982), Keel (1983) and Ho et al. (1997a). Many works have
    emphasized the aperture effect, here we just normalize the SFR by the
    corresponding aperture areas. This simple process will undoubtedly
    induce some bias, however, it is worth to see the general
    trend. Figure \ref{fig4.3.5} shows the trend of SFRs per kpc$^2$
    along the Hubble sequence. Though it is a little bitter weaker, we
    still find that early-type spirals do show higher SFRs kpc$^{-2}$
    than the late-type spirals. Thus while luminous nuclear starbursts may
    exist in the entire range of spirals (Rieke \&
    Lebofsky 1978; Devereux 1987), the relative effect is much
    stronger for the early-type spirals. We also calculated the SFRs
    from the far-Infrared emissions available in the RC3. Figure
    \ref{fig4.3.2} shows the FIR-SFRs along the Hubble types.
    Again, we could find the early-type spirals do show higher
    SFRs but not significantly exceed those of late types.

%

%

\begin{figure}
    \centering
    \includegraphics[width=8cm]{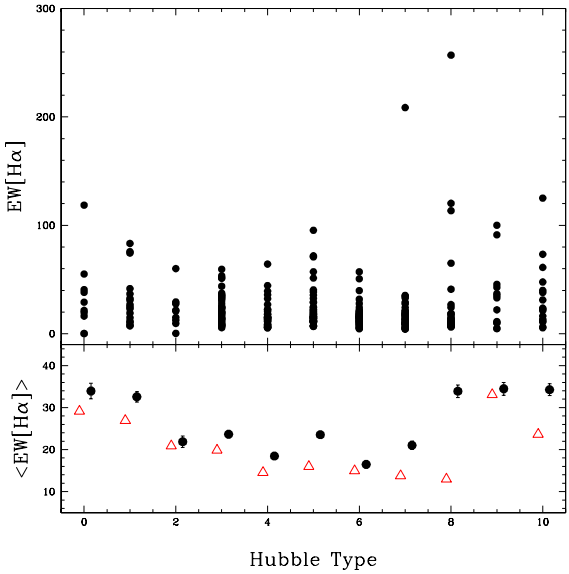}
    \caption[The distribution of EW(H$\alpha$)]{The distribution of
        EW(H$\alpha$) among the Hubble types. There is no significant
        trend found with morphology types, however, the late-type galaxies
        do show comparably larger dispersions and higher median values.
        The triangles represent the median values and the filled circles
        are the averages, which may be affected by the few extremely
        larger ones.
        \label{fig4.3.4}}
    \end{figure}

    The equivalent width of H$\alpha$ (EW[H$\alpha$]), which
    represents the efficiency of the star forming activity, shows
    the similar result (Figure \ref{fig4.2.1}). In Figure
    \ref{fig4.3.4}, we present the variations of EWs among each Hubble
    type. We find that early-type spirals show large equivalent
    widths (EWs), especially for those S0/a $\sim$ Sab galaxies.
    From Sbc to Sd type, galaxies show similar EWs, and for the
    end of the sequence (Sdm $\sim$ I) we see clearly a large
    fraction of high equivalent widths.

    For all Hubble types, the moderate values of both SFRs and EWs
    of H$\alpha$ are very common in the distributions, however,
    large dispersion among the same type do exit, which is
    consistent with James et al. (2004). Furthermore, we also find
    some elliptical and lenticular galaxies show considerable high
    SFR(H$\alpha$) and EW(H$\alpha$). Since H$\alpha$
    emission has a different origin in ellipticals and S0's than
    in spiral galaxies,  which is usually due to the nuclear
    activities or merger/peculiar interactions, we will not
    discuss these ellipticals and S0's in this paper.

\subsection{Stellar Features}
    We have measured a set of stellar indices directly from the
    synthetic spectra. For most of the stellar features,
    we use the same indices as defined by Cid Fernandes et al (2004a),
    which are based on the studies of star cluster and galaxy spectra
    by Bica \& Alloin (1986a,b) and Bica (1988). We adopt the
    definition of Worthey \& Ottaviani (1997) and Balogh et al (1999)
    for H$\delta_{A}$ and D$_{n}(4000)$ respectively. The results
    of the absorption lines are presented in Table \ref{Table4}:
    EWs of CaII K 3933 (Column 2), CN band 4200 (Column 3), G band
    4300 (Column 4), Mg Ib 5173 (Column 5), H$\delta$ 4101 (Column
    6), H$\gamma$ 4340 (Column 7), H$\beta$ 4861 (Column 8), H$\alpha$
    6563 (Column 9), and D$_{n}(4000)$ (Column 10), all in units of \AA.

    \begin{figure}
    \centering
    \includegraphics[width=8cm]{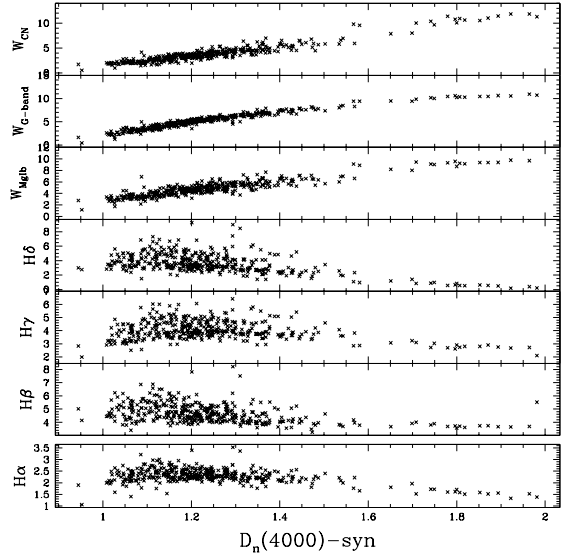}
    \caption[Relations between absorption features and
        D$_n$(4000)]{Relations between D$_{n}$(4000) and equivalent widths
        of CN band, G band, Mg Ib, H$\delta$, H$\gamma$, H$\beta$
        and H$\alpha$ measured in the synthetic spectrum. Correlations are
        clearly seen in almost all diagrams, however, the first three metal
        absorptions (top) show the strongest. \label{fig4.4.1}}
    \end{figure}
%
    \begin{figure}
    \centering
    \includegraphics[width=8cm]{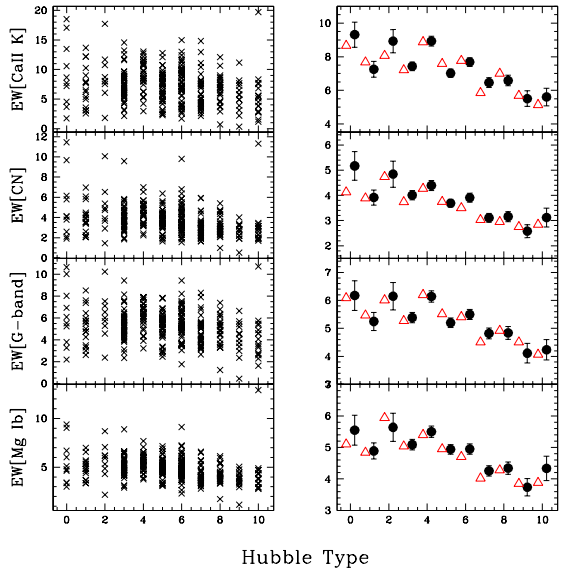}
    \caption[Relations between metal absorption lines and Hubble types]
        {Relations between Hubble types and equivalent widths
        of CaII K, CN band, G band and Mg Ib lines measured in the synthetic
        spectrum. There exist a decreasing trend from early-type spirals
        to the late-type, especially for those later than Sbc. Symbols are
        the same meaning as Figure \ref{fig4.1.2}
        \label{fig4.4.2}}
    \end{figure}
%
    \begin{figure}
    \centering
    \includegraphics[width=8cm]{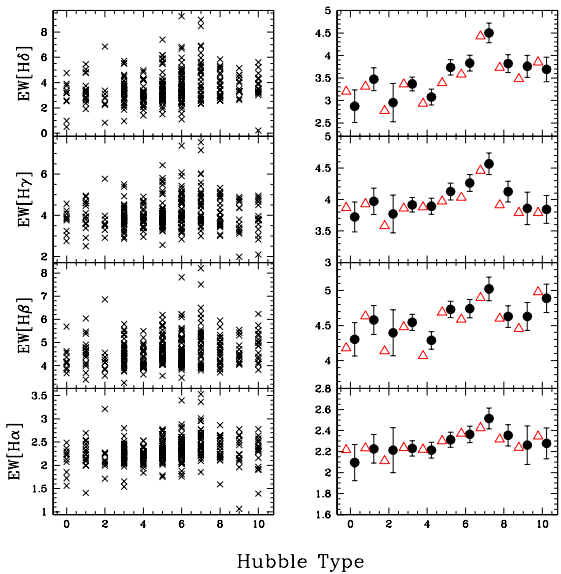}
    \caption[Relations between Balmer absorption lines and Hubble types]
        {Relations between Hubble types and equivalent widths
        H$\delta$, H$\gamma$, H$\beta$ and H$\alpha$ absorption lines measured in the
        synthetic spectrum. Interesting increasing EWs from Sbc to Sd type
        is found in all four measured Balmer absorption lines.
        \label{fig4.4.3}}
    \end{figure}

    Kauffmann et al. (2003) have found that D$_n(4000)$ is a useful
    indicator for stellar population during their study of SDSS
    galaxies, which has also been confirmed by the study of Seyfert 2
    galaxies (Cid Fernandes et al 2004b). In Figure \ref{fig4.4.1}, we
    present the relations between D$_n(4000)$ and model-derived
    equivalent widths of 3 absorption lines in the Bica's system:
    CN band, G band, Mg Ib and 3 Balmer absorption lines (H$\beta$,
    H$\gamma$ and H$\delta$). The equivalent widths of CN band, G-band
    and Mg Ib are strongly correlate with D$_n(4000)$, and the
    three Balmer lines do exit correlation, although not very tight
    comparably.

    One of the useful application of stellar population synthesis
    methods is to decompose the stellar contributions from the
    integrated spectrum to obtain pure nebular emission lines. Thus
    to derive accurate corrections of Balmer absorption features is
    particularly necessary. Previous study by McCall, Rybski \&
    Shields (1985) recommended a mean absorption correction of 2 \AA\
    to the EWs for the first three Balmer lines. Veilleux et al.
    (1995) also obtained a correction 2 \AA\ for EW(H$\beta$) of the
    old stellar population, ranging from 1 to 3 \AA. More recently, in
    the study of SFHs for a sample of starburst galaxies, Mayya et al.
    (2004) got the mean absorption EWs of 1.57, 2.48, 2.32, 2.49 \AA\
    for the first four Balmer lines. In our studies, the mean
    average Balmer absorption EWs of H$\alpha$, H$\beta$, H$\gamma$
    and H$\delta$ are 2.30, 4.68, 4.07 and 3.62, respectively.

    Figure \ref{fig4.4.2} and \ref{fig4.4.3} show the distributions of
    8 stellar absorption EWs: CaII K, CN band, G-band, Mg Ib,
    H$\alpha$, H$\beta$, H$\gamma$ and H$\delta$ along the Hubble
    types. The four metallic absorption features show a clear trend
    decreasing from the early-type sprials to late-type and irregular
    ones, especially for those later than Sbc. On the other hand, the
    four Balmer absorption lines exhibit an interesting increasing
    from Sbc to Sd type galaxies, though all these variations are
    actually small. Thus, the correction due to the Balmer absorption
    features should not be considered as a completely uniform value,
    but rather as a variation among the Hubble types.

\section{Bar Influence On the Star Forming Activity}
    Bar structures are well known to efficiently enhance the star
    formation activities of the circumnuclear regions (Huang et
    al. 1996, Ho et al. 1997b, Kennicutt 1998). Many numerical simulations
    have predict that a bar can effectively drive radial inflow of gas
    toward the center of a galaxy, which induce the burst of star
    formation (e.g., Roberts, Huntley \& van Albada 1979; Elmegreen
    1988; Athanassoula 1992; Piner, Stone \& Teuben 1995).

    \begin{center}
    \begin{table*}[htb]
    \caption[Bar frequencies]{The bar existence among different
        morphological types}
    \label{tb3}
    \begin{tabular}{ccccccccccccc}
    \hline
        Galaxy Type & L & S0/a & Sa & Sab & Sb & Sbc & Sc & Scd & Sd & Sdm & Sm & Irr \\
    \hline
        Normal & 22 & 3  & 11 & 4 & 16 & 15 & 19 & 23 & 18 & 13 &  2 & 11\\
        SAB    &  3 & 0  &  1 & 1 &  6 &  7 & 13 &  8 &  4 &  5 &  1 &  2\\
        SB     &  4 & 7  &  4 & 4 & 19 &  7 & 12 & 16 & 13 &  7 & 10 &  3\\
    \hline
    \end{tabular}
    \end{table*}
    \end{center}
    In all the 385 star forming galaxies, 157 are non-barred galaxies,
    51 are weak barred galaxies (SAB), and 106 contain strong bars
    (SB), ranging from lenticular to irregular galaxies. Thus we
    get almost equal number of barred and unbarred galaxies in our
    sample (there are another 3 barred galaxies in the remaining 65
    galaxies with uncertain morphologies, while the other 6 are elliptical
    galaxies). It is obvious that barred galaxies do show a quite
    common existence in every individual Hubble types.

    \begin{figure}
    \centering
    \includegraphics[width=8cm]{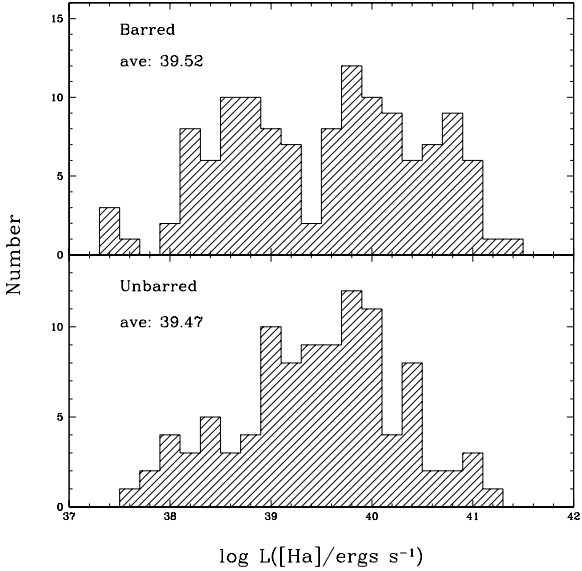}
    \caption[Distribution of extinction-corrected
        L(H$\alpha$)]{Distribution of H$\alpha$ luminosities, extinction
        corrected. From top to bottom are barred and unbarred galaxies.
        Interesting gap between barred higher L(H$\alpha$) part and lower
        one of barred galaxies also exist in Fig. 3 of Ho et al. (1997b).
        \label{fig5.1}}
    \end{figure}
%
    \begin{figure}
    \centering
    \includegraphics[width=8cm]{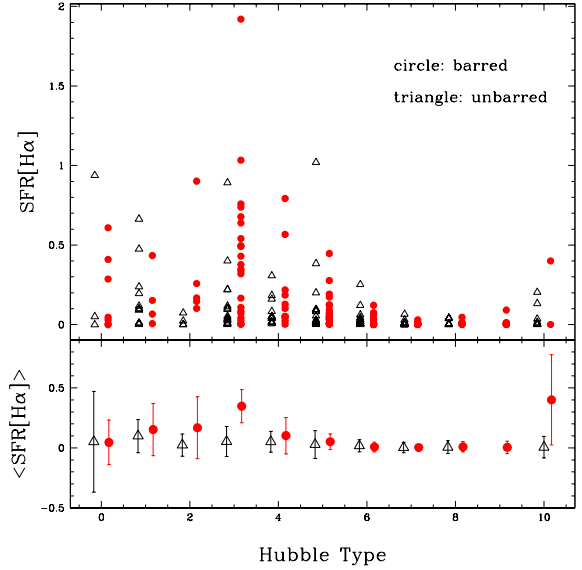}
    \caption[Comparison of SFRs between barred and unbarred
        galaxies]{The SFRs of barred and unbarred galaxies as a function
        of Hubble T-type. From top to bottom are the distribution of
        individual galaxies and their average values. Filled circles
        represent the barred galaxies, while the open triangles show the
        unbarred ones. The errorbar is the deviation in the mean.
        \label{fig5.2}}
    \end{figure}
%
    \begin{figure}
    \centering
    \includegraphics[width=8cm]{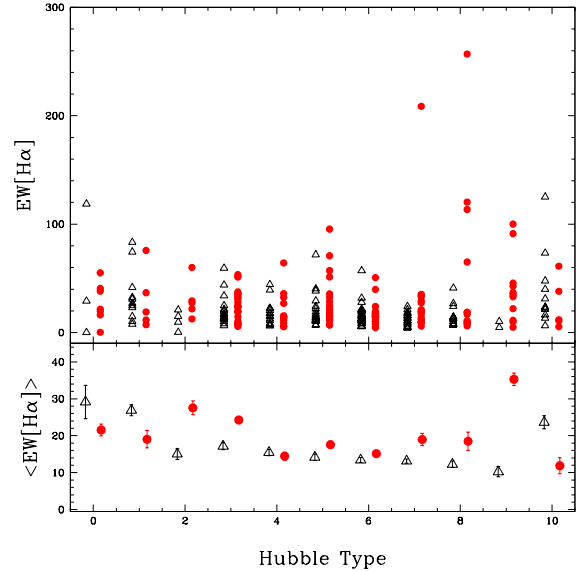}
    \caption[Distributions of EW(H$\alpha$)]{The EW[H$\alpha$] of
        barred and unbarred galaxies as a function of Hubble
        T-type. From top to bottom are the distribution of individual
        galaxies and their average values. Symbols are the same
        meaning as Figure \ref{fig5.2}
    \label{fig5.3}}
    \end{figure}
%
    \begin{figure}
    \centering
    \includegraphics[width=8cm]{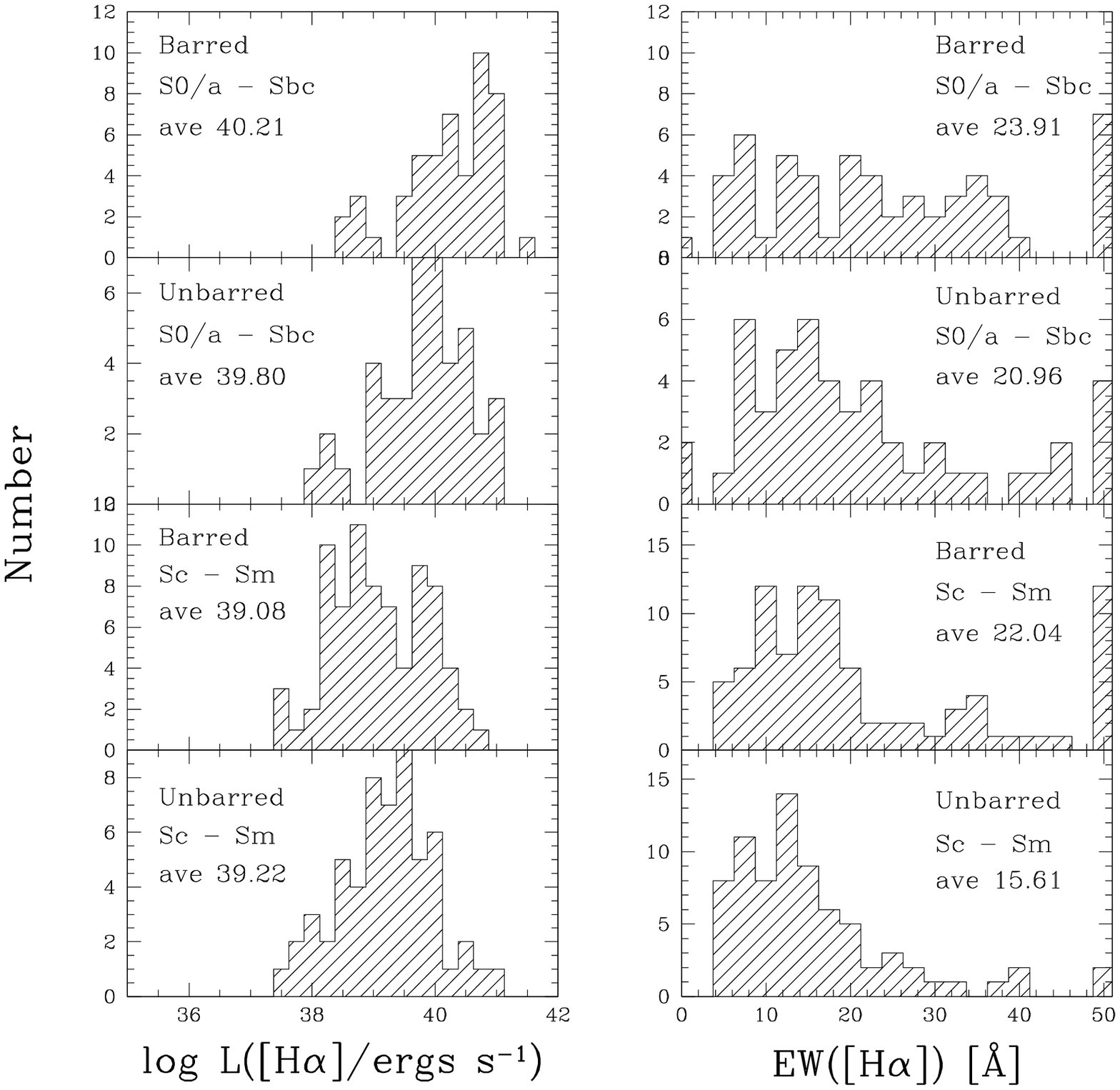}
    \caption[Distributions of extinction-corrected L(H$\alpha$) and
    EW(H$\alpha$)] {The histograms of the extinction-corrected
    L[H$\alpha$] (left) and EW(H$\alpha$)s (right). The two top panels
    show the barred and unbarred galaxies in early-type spirals
    (S0/a-Sbc) and the two bottoms show the late-type spirals (Sc-Sm).
    \label{fig5.4}}
    \end{figure}

    The comparison between H$\alpha$ luminosities in barred and
    unbarred galaxies is shown in Figure \ref{fig5.1}. The barred
    galaxies show a large fraction toward the high end of
    L(H$\alpha$). Interesting gap between high and low part of
    luminosities does also exist in Fig. 3 of Ho et al. (1997b).
    Figure \ref{fig5.2} shows the distribution of
    both the individual galaxies and the mean star formation rates of
    our sample, but this time separated into barred (filled circle)
    and unbarred (open triangle) types. It is clear that bars
    can enhance star formation, or at
    least extend the range toward the higher SFRs. This trend mainly
    occurs in the early-type spirals which is consistent
    with previous study of Huang et al. (1996) and Ho et al. (1997b).

    Less agreement is found in the distribution of EW(H$\alpha$)s
    (Figure \ref{fig5.3}). In general, the equivalent widths of H$\alpha$
    of barred galaxies are higher than unbarred ones, but not
    significant in many individual types. We do find clear
    enhancement in EW(H$\alpha$) of barred galaxies (Figure
    \ref{fig5.4}), however, the largest difference is seen in the
    very late types. Since late-type spirals are believed to be gas
    rich, thus, other influence like gas contents and interactions may
    also play an important roll in the circulmnuclear star formation
    activities.

\section{Conclusions}
    We have constructed a sample of galaxies by positionally matching
    up the SDSS and RC3 data. We then applied the stellar population
    synthesis model to statistically study the variations of the star
    formation histories and line features along the Hubble types. Our
    finding are summarized as follows:
   \begin{enumerate}
      \item A decreasing sequence is found for nebular extinction from
        early-type spirals to late-type ones, which is consistent with
        previous studies of Ho et al. (1997a) and Stasi\'nska et al. (2004).
        We have also confirmed that nebular emissions do suffer more
        serious extinctions than stellar light and the later one shows
        almost no variations among the Hubble types.
      \item We find that H$\alpha$ luminosities of early-type spiral
        galaxies are much higher than the late-type ones, which causes the
        corresponding larger SFRs in early-type spirals.
        From the equivalent widths of H$\alpha$, we find in
        general, early-type spirals show higher value of EWs, however, the
        dramatic change is found from Sdm type. We also plot the FIR-SFRs
        as a function of Hubble types and found that the early-type
        spirals show higher amount of FIR-SFRs and may partly due to the dust
        re-emission which will favor the early types. However, this
        difference of SFRs in IR band is comparably much smaller.
        Although large amount of galaxies show quite modest SFRs and may
        not actually contain bright HII regions, those who have strongly
        favor the higher star-forming activities in early-type galaxies.
      \item We have measured the spectral absorption features and
        confirmed previous findings that tight correlations between
        D$_n$(4000) and equivalent widths of absorption lines: CN band,
        G-band, Mg Ib, H$\delta$, H$\gamma$, H$\beta$ and H$\alpha$.
        The dependence of mean EWs on the Hubble types could be seen in
        metallic absorption lines. We also find an increase sequence of
        mean Balmer absorption equivalent widths from Sbc to Sd type.
        Since all these EWs are measured automatically, they do
        not depend on the morphology types.
      \item We have discussed the bar effects on the star forming
        activities on both early-type and late-type spirals, and confirmed
        previous findings that bar structures can enhance the star forming
        activities, especially in early type spiral galaxies. However,
        bars are not necessarily inducing strong star formation activities;
        and those without could still contain high star formation regions.
        We also find that, though in general, the barred early-type
        spirals show larger EW[H$\alpha$], the largest difference of
        EW[H$\alpha$] between barred and unbarred galaxies do exist in the
        very late-type spirals (Sd $\sim$ Sm). Thus we believe that other
        effects like interactions and gas contents may also play important
        roles in the star formation activities.
   \end{enumerate}

\begin{acknowledgements}
 We would like to thank the anonymous referee for instructive
 comments which improved the content of the paper and thank
 Roberto Cid Fernandes for sending us the updated code for stellar
 population synthesis. This work is supported by the National
 Natural Science Foundation of China under grant 10221001 and the
 National Key Basic Research Science Foundation (NKBRSG19990754).
 Funding for the creation and distribution of the SDSS Archive has
 been provided by the Alfred P. Sloan Foundation, the
 Participating Institutions, the National Aeronautics and Space
 Administration, the National Science Foundation, the U.S.
 Department of Energy, the Japanese Monbukagakusho, and the Max
 Planck Society. The SDSS web site is http://www.sdss.org/.  The
 SDSS is managed by the Astrophysical Research Consortium (ARC)
 for the Participating Institutions. The Participating
 Institutions are The University of Chicago, Fermilab, the
 Institute for Advanced Study, the Japan Participation Group, The
 Johns Hopkins University, Los Alamos National Laboratory, the
 Max-Planck-Institute for Astronomy (MPIA), the
 Max-Planck-Institute for Astrophysics (MPA), New Mexico State
 University, University of Pittsburgh, Princeton University, the
 United States Naval Observatory, and the University of
 Washington. This research has made use of the NASA/IPAC
 Extragalactic Database (NED) which is operated by the Jet
 Propulsion Laboratory, California Institute of Technology, under
 contract with the National Aeronautics and Space Administration.
\end{acknowledgements}

\begin{longtable}{lllccccl}
\caption{\label{Table1} Catalog of the whole sample}\\
\hline\hline
    Name         &  R.A.(2000) &  Dec.(2000)  & Redshift & Av$_{st}$& Av$_{neb}$& log L$_{FIR}^a$& Morphology\\
\hline
\endfirsthead
\caption{continued.}\\
\hline\hline
    Name         &  R.A.(2000) &  Dec.(2000)  & Redshift & Av$_{st}$& Av$_{neb}$& log L$_{FIR}^a$& Morphology\\
\hline
\endhead
\hline
\endfoot
NGC3042          & 09h53m20.2s & +00d41m51.8s & 0.012613 &   0.1073 &   - - -  &   - -  & S0           \\
UGC5238          & 09h46m53.5s & +00d30m26.4s & 0.005929 &   0.7515 &   0.7775 &   - -  & SBdm         \\
UGC7963          & 12h47m52.9s & -01d11m08.9s & 0.023371 &   0.9519 &   - - -  &   - -  & Sdm          \\
A1252+00         & 12h55m12.6s & +00d06m59.9s & 0.004180 &   0.0295 &   0.5290 &   - -  & SBd          \\
NGC4668          & 12h45m31.9s & -00d32m08.5s & 0.005431 &   0.1912 &   0.3029 &  42.38 & SBd          \\
CGCG15-20        & 12h47m19.3s & +00d24m17.4s & 0.047169 &  -0.0587 &   - - -  &   - -  & S0           \\
MCG0-29-27       & 11h24m08.6s & -01d09m27.9s & 0.029288 &   0.5686 &   1.0296 &   - -  & SAB0         \\
UGC6435          & 11h25m35.0s & -00d46m05.6s & 0.025325 &  -0.0924 &   - - -  &   - -  & S0           \\
UGC6340          & 11h19m55.4s & -00d52m47.9s & 0.024573 &  -0.0535 &   - - -  &  43.65 & SABbc        \\
UGC6432          & 11h25m17.9s & +00d21m01.8s & 0.040292 &   0.2952 &   0.2812 &   - -  & SAbc         \\
CGCG11-100       & 11h27m36.7s & +00d23m42.8s & 0.049075 &   0.3285 &   - - -  &   - -  & SBb          \\
MCG0-29-29       & 11h24m18.6s & +00d38m37.4s & 0.026402 &   0.2049 &   1.3175 &   - -  & SBc          \\
UGC9299          & 14h29m34.6s & -00d01m05.7s & 0.005204 &   0.2909 &  -0.1752 &   - -  & SABd         \\
IC1010           & 14h27m20.3s & +01d01m33.2s & 0.025687 &   0.2471 &   - - -  &   - -  & SBb          \\
MCG0-29-36       & 11h28m16.4s & +00d53m28.9s & 0.039884 &   0.2740 &   0.7626 &   - -  & SABbc        \\
UGC6402          & 11h23m19.1s & -00d55m21.4s & 0.008698 &   0.9888 &   1.3155 &  43.04 & Sdm          \\
UGC6457          & 11h27m12.2s & -00d59m40.8s & 0.003191 &   0.5764 &  -0.0508 &   - -  & dIn          \\
MCG0-30-4        & 11h31m58.6s & -00d03m01.1s & 0.039859 &   0.7349 &   - - -  &   - -  & SAcd         \\
MCG0-29-28       & 11h24m09.1s & +00d42m01.9s & 0.025987 &   0.4158 &   0.0219 &   - -  & Sc           \\
NGC3719          & 11h32m13.4s & +00d49m09.3s & 0.019457 &   0.1252 &   0.3016 &   - -  & SAbc         \\
CGCG11-103       & 11h27m57.5s & -01d12m40.1s & 0.042687 &   1.5888 &   1.9545 &   - -  & SA0          \\
IC992            & 14h18m14.9s & +00d53m28.0s & 0.025965 &   0.3778 &   1.1523 &  43.75 & SABc         \\
UGC5195          & 09h43m12.0s & +00d24m51.0s & 0.025190 &   0.8237 &   0.9703 &  43.74 & Sbc          \\
UGC5205          & 09h44m07.2s & -00d39m29.7s & 0.005003 &  -1.6866 &   - - -  &   - -  & SBm          \\
UGC5242          & 09h47m05.5s & +00d57m51.9s & 0.006137 &  -0.1979 &   0.0223 &   - -  & SBm          \\
IC1011           & 14h28m04.5s & +01d00m22.8s & 0.025647 &   0.3966 &   0.6289 &  43.68 & ----         \\
UGC6608          & 11h38m33.2s & -01d11m04.2s & 0.020776 &   0.8511 &   1.4847 &  43.52 & SABab        \\
MCG0-30-7        & 11h32m45.4s & -00d44m27.7s & 0.022379 &   0.4967 &   0.2987 &   - -  & SBm          \\
CGCG12-54        & 11h40m06.4s & -00d50m15.7s & 0.019591 &  -0.1357 &   - - -  &   - -  & E            \\
CGCG12-27        & 11h34m39.1s & +00d07m29.1s & 0.028791 &   0.2434 &   - - -  &   - -  & SA0/a        \\
CGCG12-5         & 11h32m09.2s & -00d56m33.4s & 0.026154 &  -0.0494 &   - - -  &   - -  & SAB0         \\
CGCG12-53        & 11h40m06.2s & -00d54m05.0s & 0.028712 &   0.6698 &   1.5266 &   - -  & SAb          \\
IC716            & 11h39m03.3s & -00d12m21.6s & 0.018078 &   0.6377 &   - - -  &   - -  & Sbc          \\
NGC3720          & 11h32m21.8s & +00d48m17.0s & 0.020068 &   0.1563 &   0.9369 &  44.09 & SAa          \\
NGC5750          & 14h46m11.1s & -00d13m22.6s & 0.005827 &   0.3941 &   - - -  &  42.52 & SB0/a        \\
NGC5733          & 14h42m45.9s & -00d21m03.8s & 0.005698 &   0.4660 &   0.0989 &   - -  & dIn          \\
UGC9470          & 14h41m48.6s & +00d41m13.1s & 0.006427 &   1.3992 &   - - -  &   - -  & SBdm         \\
UGC9977          & 15h41m59.5s & +00d42m46.0s & 0.006483 &   1.3629 &   0.1903 &  42.41 & Sc           \\
MCG0-39-4        & 15h07m59.2s & +01d13m54.2s & 0.035114 &   0.5962 &   1.5006 &  44.05 & S?           \\
UGC9732          & 15h08m09.7s & +01d14m57.6s & 0.035335 &   0.4407 &   0.7379 &   - -  & SBb          \\
CGCG17-54        & 13h39m13.2s & -01d07m15.4s & 0.014852 &   0.2651 &   0.3066 &   - -  & Sb           \\
CGCG17-43        & 13h38m06.4s & +00d01m13.7s & 0.022360 &  -0.1490 &   0.2262 &   - -  & HII?         \\
NGC5887          & 15h14m43.9s & +01d09m15.4s & 0.029286 &   0.1289 &   - - -  &   - -  & SA0          \\
UGC10264         & 16h12m56.8s & -00d05m46.5s & 0.030845 &   0.6339 &   - - -  &  43.88 & SABcd        \\
MCG0-41-5        & 16h10m14.6s & +01d03m20.5s & 0.027844 &   0.6427 &   - - -  &   - -  & SABdm        \\
IC4229           & 13h22m26.1s & -02d25m05.7s & 0.023181 &   0.2678 &   0.9835 &  43.83 & SBb          \\
UGC10005         & 15h45m14.3s & +00d46m19.9s & 0.012830 &   0.3854 &   - - -  &   - -  & SAd          \\
A1301-03         & 13h04m31.1s & -03d34m20.6s & 0.004623 &  -0.2095 &   0.1498 &   - -  & SABdm        \\
UGC10306         & 16h16m43.5s & +00d14m47.2s & 0.030720 &   0.0906 &   - - -  &   - -  & SABb         \\
MK502            & 16h53m42.9s & +64d05m06.8s & 0.041539 &   0.1778 &   - - -  &   - -  & Compact      \\
IC1235           & 16h52m03.6s & +63d06m56.8s & 0.010433 &   0.1460 &   0.0481 &   - -  & ----         \\
IC1248           & 17h11m40.1s & +59d59m44.2s & 0.016699 &   0.3598 &   0.7844 &   - -  & SBc          \\
\end{longtable}
\noindent $^a$ Luminosity in unit of erg s$^{-1}$

\begin{longtable}{llllllll}
\caption{\label{Table2} Flux densities of emissions}\\
\hline\hline
    Name        &   F4861 &   F5007 &   F6548 &   F6584 &   F6563 &   F6717 &   F6731\\
\hline
\endfirsthead
\caption{continued.}\\
\hline\hline
    Name        &   F4861 &   F5007 &   F6548 &   F6584 &   F6563 &   F6717 &   F6731\\
\hline
\endhead
\hline
\endfoot
NGC3042         &   87.60 &  206.83 &  215.99 &  715.24 &  356.47 &  173.07 &  174.66\\
UGC5238         &  221.14 &  343.76 &   50.40 &  131.92 &  839.97 &  136.81 &  100.51\\
UGC7963         &   40.52 &   16.64 &   25.60 &   68.87 &  169.61 &   35.46 &   25.17\\
A1252+00        &  117.13 &   69.67 &   61.47 &  133.81 &  406.33 &  106.49 &   75.72\\
NGC4668         &   77.33 &  111.44 &   25.37 &   62.83 &  247.01 &   50.40 &   40.74\\
CGCG15-20       &   32.83 &   63.76 &   41.08 &  100.76 &   85.81 &   33.22 &   37.83\\
MCG0-29-27      &  454.87 & 3623.94 &  474.27 & 1397.25 & 1894.18 &  476.63 &  436.90\\
UGC6435         &   32.27 &   54.61 &   54.48 &  143.85 &  100.85 &   45.46 &   63.48\\
UGC6340         &   59.64 &  113.27 &   52.29 &  133.68 &  159.19 &   93.57 &   83.84\\
UGC6432         &   67.03 &   40.45 &   27.58 &   69.79 &  212.42 &   29.59 &   23.91\\
CGCG11-100      &   84.60 &   59.61 &   48.37 &  170.48 &  382.18 &   53.53 &   41.48\\
MCG0-29-29      &  205.59 &   60.20 &  100.57 &  340.95 &  950.97 &  155.27 &  115.74\\
UGC9299         &  100.17 &  142.39 &   19.46 &   49.81 &  268.74 &   60.93 &   41.82\\
IC1010          &   28.23 &   50.21 &   37.40 &  114.03 &   77.11 &   40.71 &   45.17\\
MCG0-29-36      &  118.48 &  100.65 &   71.61 &  169.99 &  447.58 &   43.43 &   33.22\\
UGC6402         &  262.24 &  321.97 &   90.41 &  255.23 & 1212.12 &  240.04 &  170.32\\
UGC6457         &   74.76 &  119.94 &    6.10 &   19.30 &  209.89 &   47.78 &   35.90\\
MCG0-30-4       &   24.88 &   10.97 &    7.94 &   23.57 &   53.15 &    7.72 &    7.94\\
MCG0-29-28      &   85.01 &   26.07 &   32.01 &   90.03 &  245.07 &   40.70 &   28.08\\
NGC3719         &   99.62 &   73.50 &   69.68 &  206.05 &  318.08 &   79.54 &   64.01\\
CGCG11-103      &  176.13 &  597.52 &  371.61 & 1052.18 & 1027.92 &  186.33 &  192.54\\
IC992           &  413.55 &  149.43 &  239.34 &  733.45 & 1800.94 &  195.61 &  155.79\\
UGC5195         &   75.89 &   50.32 &   56.59 &  174.65 &  309.25 &   65.11 &   45.78\\
UGC5205         &    0.88 &    4.33 &    3.68 &   11.91 &    8.45 &    1.88 &    3.18\\
UGC5242         &   73.65 &  142.13 &   11.32 &   21.61 &  212.37 &   33.74 &   24.61\\
IC1011          &  264.94 &  116.42 &  133.05 &  427.08 &  953.21 &  174.78 &  131.63\\
UGC6608         &  144.86 &   51.48 &  115.72 &  339.64 &  712.22 &  116.92 &  101.07\\
MCG0-30-7       &   41.99 &   31.64 &   12.77 &   30.72 &  133.92 &   42.33 &   27.34\\
CGCG12-54       &   13.71 &   58.07 &   24.72 &   43.51 &   43.81 &    4.70 &   19.61\\
CGCG12-27       &   14.47 &   11.71 &    8.74 &   15.70 &   20.60 &    7.53 &    3.79\\
CGCG12-5        &   65.53 &  184.96 &  126.14 &  266.62 &  220.74 &   91.16 &   82.65\\
CGCG12-53       &  198.49 &   40.59 &  113.06 &  327.10 &  990.92 &  123.50 &   93.72\\
IC716           &   47.00 &   57.84 &   32.01 &  169.37 &  159.05 &   64.46 &   63.97\\
NGC3720         &  180.55 &   38.49 &   78.30 &  233.36 &  726.85 &   66.07 &   53.76\\
NGC5750         &  216.90 &  331.13 &  306.15 &  869.79 &  802.45 &  525.02 &  399.84\\
NGC5733         &  223.31 &  505.47 &   27.75 &   83.83 &  662.12 &  132.27 &   88.91\\
UGC9470         &   17.79 &   24.47 &   12.69 &   12.06 &   52.33 &   14.95 &   12.16\\
UGC9977         &   51.81 &   29.84 &   16.95 &   49.62 &  158.83 &   48.37 &   33.27\\
MCG0-39-4       &  855.21 &  292.04 &  764.47 & 2451.81 & 4229.11 &  752.21 &  658.27\\
UGC9732         &   81.72 &   90.88 &   98.14 &  250.93 &  305.94 &  108.88 &   77.93\\
CGCG17-54       &   78.90 &   28.70 &   24.87 &   86.00 &  252.37 &   57.94 &   38.02\\
CGCG17-43       &  153.46 &  136.86 &   38.97 &  125.39 &  476.67 &  110.59 &   79.35\\
NGC5887         &   64.82 &  110.90 &  122.43 &  299.64 &  182.45 &  124.74 &  108.50\\
UGC10264        &   30.11 &   87.02 &   22.06 &   77.12 &   94.72 &   32.00 &   21.27\\
MCG0-41-5       &   20.38 &   31.95 &   14.14 &   34.65 &   32.84 &   13.35 &   14.66\\
IC4229          & 1039.08 &  189.08 &  562.62 & 1755.09 & 4254.76 &  538.92 &  442.83\\
UGC10005        &   27.08 &   11.16 &    9.99 &   22.88 &   52.19 &   15.34 &   12.29\\
A1301-03        &  296.09 &  671.88 &   30.27 &   93.71 &  894.39 &  130.47 &   91.69\\
UGC10306        &   51.46 &   92.14 &   89.49 &  254.92 &  173.67 &  103.60 &   83.59\\
MK502           &   23.63 &   13.44 &   11.75 &   32.05 &   51.36 &   10.20 &    7.92\\
IC1235          &  233.19 &  210.91 &   60.58 &  170.31 &  678.74 &  180.12 &  126.67\\
IC1248          &  176.12 &   43.94 &   68.81 &  216.49 &  670.64 &  115.54 &   76.33\\
\end{longtable}
\noindent Flux densities are in units of $10^{-17} \times erg\ s^{-1} cm^{-2}$\AA.\\

\begin{longtable}{lccccccc}
\caption{\label{Table3} Equivalent widths of Emissions}\\
\hline\hline
   Name         &  EW4861  &   EW5007 &   EW6548 &   EW6584 &   EW6563 &   EW6717 &  EW6731\\
\hline
\endfirsthead
\caption{continued.}\\
\hline\hline
   Name         &  EW4861  &   EW5007 &   EW6548 &   EW6584 &   EW6563 &   EW6717 &  EW6731\\
\hline
\endhead
\hline
\endfoot
NGC3042         &    0.31 &    0.67 &    0.71 &    2.33 &    1.17 &    0.52 &    0.53\\
UGC5238         &   34.32 &   43.76 &    7.29 &   18.64 &  120.30 &   21.19 &   15.58\\
UGC7963         &    3.52 &    1.30 &    2.14 &    5.69 &   14.12 &    2.54 &    1.80\\
A1252+00        &    3.73 &    1.91 &    1.87 &    3.99 &   12.25 &    4.05 &    2.89\\
NGC4668         &    3.18 &    3.90 &    0.94 &    2.32 &    9.14 &    2.24 &    1.81\\
CGCG15-20       &    0.50 &    0.92 &    0.64 &    1.56 &    1.33 &    0.47 &    0.54\\
MCG0-29-27      &    6.72 &   45.88 &    6.41 &   18.07 &   25.12 &    7.14 &    6.55\\
UGC6435         &    0.17 &    0.26 &    0.28 &    0.75 &    0.52 &    0.22 &    0.31\\
UGC6340         &    0.39 &    0.70 &    0.33 &    0.86 &    1.02 &    0.58 &    0.52\\
UGC6432         &    2.54 &    1.34 &    0.96 &    2.37 &    7.32 &    1.08 &    0.87\\
CGCG11-100      &    1.92 &    1.29 &    1.04 &    3.66 &    8.22 &    1.20 &    0.93\\
MCG0-29-29      &    6.48 &    1.77 &    3.02 &   10.27 &   28.62 &    6.24 &    4.66\\
UGC9299         &    8.85 &    9.53 &    1.41 &    3.57 &   19.39 &    5.66 &    3.89\\
IC1010          &    0.54 &    0.93 &    0.69 &    2.11 &    1.42 &    0.69 &    0.77\\
MCG0-29-36      &    3.17 &    2.50 &    1.82 &    4.27 &   11.31 &    1.20 &    0.92\\
UGC6402         &   10.34 &   10.50 &    3.10 &    8.52 &   41.06 &    9.12 &    6.47\\
UGC6457         &    6.87 &   11.27 &    0.59 &    1.87 &   20.27 &    6.77 &    5.09\\
MCG0-30-4       &    3.84 &    1.51 &    1.10 &    3.23 &    7.32 &    1.11 &    1.14\\
MCG0-29-28      &    4.34 &    1.16 &    1.51 &    4.14 &   11.41 &    2.00 &    1.38\\
NGC3719         &    2.36 &    1.65 &    1.61 &    4.74 &    7.33 &    1.89 &    1.52\\
CGCG11-103      &    6.78 &   18.08 &   10.40 &   29.90 &   28.96 &    4.02 &    4.15\\
IC992           &   12.18 &    4.12 &    6.86 &   20.75 &   51.35 &    6.75 &    5.38\\
UGC5195         &    3.03 &    1.73 &    2.04 &    6.38 &   11.19 &    2.08 &    1.46\\
UGC5205         &    0.12 &    0.43 &    0.41 &    1.40 &    0.97 &    0.36 &    0.62\\
UGC5242         &   15.28 &   27.75 &    2.50 &    4.46 &   45.58 &    9.73 &    7.11\\
IC1011          &    5.85 &    2.30 &    2.71 &    8.80 &   19.53 &    4.01 &    3.02\\
UGC6608         &    4.85 &    1.58 &    3.56 &   10.33 &   21.78 &    3.25 &    2.81\\
MCG0-30-7       &    7.33 &    5.35 &    2.14 &    4.95 &   22.07 &    8.23 &    5.32\\
CGCG12-54       &    0.16 &    0.68 &    0.30 &    0.53 &    0.53 &    0.05 &    0.23\\
CGCG12-27       &    1.37 &    1.00 &    0.76 &    1.36 &    1.80 &    0.65 &    0.33\\
CGCG12-5        &    0.72 &    1.87 &    1.39 &    2.89 &    2.41 &    0.95 &    0.87\\
CGCG12-53       &    8.12 &    1.36 &    3.89 &   11.10 &   33.90 &    4.49 &    3.41\\
IC716           &    0.76 &    0.86 &    0.47 &    2.58 &    2.38 &    0.82 &    0.81\\
NGC3720         &    6.26 &    1.18 &    2.49 &    7.38 &   23.08 &    2.70 &    2.20\\
NGC5750         &    1.21 &    1.68 &    1.62 &    4.59 &    4.24 &    2.59 &    1.98\\
NGC5733         &    7.76 &   16.69 &    0.95 &    2.85 &   22.59 &    6.07 &    4.09\\
UGC9470         &    3.04 &    4.15 &    2.07 &    1.91 &    8.42 &    2.89 &    2.35\\
UGC9977         &    4.18 &    2.33 &    1.25 &    3.75 &   11.81 &    3.26 &    2.24\\
MCG0-39-4       &   11.36 &    3.61 &    9.36 &   30.28 &   51.96 &    9.85 &    8.62\\
UGC9732         &    1.41 &    1.38 &    1.60 &    4.11 &    5.00 &    1.60 &    1.14\\
CGCG17-54       &    5.97 &    1.78 &    1.69 &    5.95 &   17.28 &    5.03 &    3.30\\
CGCG17-43       &   10.58 &    7.40 &    2.15 &    6.93 &   26.32 &    8.33 &    5.99\\
NGC5887         &    0.48 &    0.71 &    0.84 &    2.04 &    1.25 &    0.77 &    0.67\\
UGC10264        &    0.75 &    2.07 &    0.51 &    1.80 &    2.21 &    0.67 &    0.45\\
MCG0-41-5       &    0.69 &    1.01 &    0.46 &    1.13 &    1.07 &    0.36 &    0.40\\
IC4229          &    7.02 &    1.22 &    3.65 &   11.39 &   27.59 &    4.24 &    3.49\\
UGC10005        &    2.90 &    0.98 &    0.93 &    2.08 &    4.80 &    1.62 &    1.30\\
A1301-03        &   43.31 &   77.93 &    3.82 &   11.99 &  113.47 &   27.64 &   19.48\\
UGC10306        &    0.53 &    0.87 &    0.90 &    2.54 &    1.74 &    0.95 &    0.77\\
MK502           &    1.15 &    0.58 &    0.52 &    1.42 &    2.28 &    0.47 &    0.36\\
IC1235          &    5.92 &    4.50 &    1.34 &    3.76 &   14.99 &    5.41 &    3.81\\
IC1248          &    5.88 &    1.31 &    2.15 &    6.78 &   20.99 &    4.20 &    2.78\\
\end{longtable}
\noindent All equivalent wavelengths are in units of \AA.

\begin{longtable}{lccccccccc}
\caption{\label{Table4} Equivalent widths of absorptions}\\
\hline\hline
    Name        & EW3933 & EW4200 & EW4340 & EW5173 & EW4101 & EW4300 & EW4861 & EW6563 & D$_{n}(4000)$\\
\hline
\endfirsthead
\caption{continued.}\\
\hline\hline
    Name        & EW3933 & EW4200 & EW4340 & EW5173 & EW4101 & EW4300 & EW4861 & EW6563 & D$_{n}(4000)$\\
\hline
\endhead
\hline
\endfoot
NGC3042         &  17.63 &  11.38 &  10.38 &   9.19 &   0.70 &   2.96 &   3.84 &   1.64 &   1.83\\
UGC5238         &   5.66 &   2.59 &   4.41 &   3.81 &   3.06 &   3.49 &   3.94 &   2.23 &   1.15\\
UGC7963         &   8.56 &   4.73 &   6.29 &   5.95 &   2.92 &   3.84 &   4.13 &   2.11 &   1.29\\
A1252+00        &   7.50 &   3.24 &   5.42 &   4.71 &   4.07 &   4.43 &   4.78 &   2.35 &   1.20\\
NGC4668         &   7.93 &   3.26 &   5.36 &   4.58 &   3.40 &   3.98 &   4.59 &   2.41 &   1.22\\
CGCG15-20       &  17.81 &  10.89 &  10.41 &   9.36 &   0.59 &   2.96 &   3.72 &   1.49 &   1.87\\
MCG0-29-27      &   9.33 &   4.63 &   6.46 &   5.86 &   2.47 &   3.57 &   3.95 &   2.10 &   1.29\\
UGC6435         &  18.47 &  11.45 &  10.59 &   9.27 &   0.39 &   2.77 &   3.57 &   1.49 &   1.91\\
UGC6340         &  18.07 &  10.64 &  10.18 &   8.41 &   0.55 &   2.90 &   3.42 &   1.51 &   1.86\\
UGC6432         &   9.82 &   5.16 &   6.77 &   6.58 &   2.65 &   3.75 &   4.36 &   2.12 &   1.31\\
CGCG11-100      &   9.62 &   4.82 &   6.63 &   6.14 &   2.81 &   3.82 &   4.29 &   2.09 &   1.33\\
MCG0-29-29      &   2.82 &   1.84 &   3.38 &   3.31 &   4.81 &   4.05 &   4.71 &   1.75 &   1.12\\
UGC9299         &   4.14 &   2.23 &   3.99 &   3.69 &   5.54 &   5.03 &   5.40 &   2.63 &   1.14\\
IC1010          &  17.61 &   9.42 &  10.16 &   8.52 &   1.02 &   3.08 &   3.62 &   1.79 &   1.79\\
MCG0-29-36      &  11.26 &   5.45 &   7.02 &   6.37 &   2.54 &   3.70 &   4.44 &   2.23 &   1.37\\
UGC6402         &   2.90 &   2.05 &   3.31 &   3.43 &   5.84 &   5.15 &   5.67 &   2.64 &   1.13\\
UGC6457         &   2.11 &   1.81 &   2.22 &   2.62 &   3.61 &   3.35 &   4.87 &   2.27 &   1.02\\
MCG0-30-4       &   6.21 &   3.39 &   5.31 &   4.75 &   6.15 &   5.71 &   5.68 &   2.68 &   1.29\\
MCG0-29-28      &   9.49 &   4.27 &   6.35 &   5.48 &   3.32 &   4.09 &   4.40 &   2.30 &   1.31\\
NGC3719         &  13.46 &   6.50 &   8.46 &   7.39 &   1.52 &   3.12 &   3.61 &   1.89 &   1.48\\
CGCG11-103      &  13.39 &   6.54 &   7.12 &   5.10 &   3.74 &   4.32 &   4.79 &   0.90 &   1.57\\
IC992           &   5.67 &   3.18 &   4.50 &   4.55 &   3.51 &   3.84 &   4.73 &   2.29 &   1.14\\
UGC5195         &  12.53 &   5.44 &   7.52 &   5.99 &   3.00 &   4.17 &   4.29 &   2.29 &   1.48\\
UGC5205         &   8.57 &   4.92 &   6.47 &   7.45 &   2.61 &   3.40 &   5.07 &   1.70 &   1.21\\
UGC5242         &   5.82 &   2.21 &   4.53 &   3.77 &   5.14 &   4.89 &   5.54 &   2.78 &   1.16\\
IC1011          &   7.54 &   4.16 &   5.39 &   5.36 &   3.22 &   3.87 &   4.71 &   2.27 &   1.22\\
UGC6608         &  10.75 &   5.37 &   7.11 &   6.21 &   2.24 &   3.50 &   3.82 &   2.11 &   1.37\\
MCG0-30-7       &   6.50 &   3.34 &   4.58 &   4.07 &   4.62 &   4.66 &   5.30 &   2.66 &   1.20\\
CGCG12-54       &  17.88 &  11.54 &  10.65 &   9.74 &   0.62 &   2.73 &   3.86 &   1.61 &   1.84\\
CGCG12-27       &  14.44 &   6.79 &   8.44 &   7.12 &   2.02 &   3.67 &   3.98 &   2.13 &   1.55\\
CGCG12-5        &  18.00 &  10.54 &  10.54 &   9.41 &   0.57 &   2.82 &   3.62 &   1.55 &   1.84\\
CGCG12-53       &   7.22 &   4.08 &   5.21 &   4.89 &   3.39 &   4.00 &   4.53 &   2.32 &   1.21\\
IC716           &  17.04 &   9.28 &   9.89 &   8.15 &   1.24 &   3.33 &   3.69 &   1.88 &   1.77\\
NGC3720         &   4.97 &   3.47 &   4.25 &   5.10 &   2.86 &   3.38 &   4.66 &   2.09 &   1.09\\
NGC5750         &  14.87 &   8.14 &   8.75 &   7.64 &   1.94 &   3.78 &   4.21 &   2.05 &   1.60\\
NGC5733         &   3.82 &   2.29 &   3.31 &   3.32 &   3.85 &   3.84 &   4.86 &   2.42 &   1.08\\
UGC9470         &   0.70 &   0.98 &   1.24 &   1.74 &   3.78 &   3.00 &   4.08 &   1.59 &   1.03\\
UGC9977         &   4.89 &   3.35 &   4.95 &   4.35 &   7.40 &   6.43 &   6.19 &   2.78 &   1.29\\
MCG0-39-4       &   7.53 &   4.29 &   5.67 &   5.64 &   2.98 &   3.79 &   4.30 &   2.09 &   1.23\\
UGC9732         &  14.45 &   8.17 &   8.85 &   7.79 &   1.73 &   3.56 &   4.04 &   1.87 &   1.59\\
CGCG17-54       &   5.70 &   2.74 &   4.32 &   4.16 &   2.77 &   3.31 &   4.18 &   2.24 &   1.11\\
CGCG17-43       &   7.81 &   2.90 &   5.51 &   4.46 &   4.54 &   4.65 &   4.99 &   2.55 &   1.24\\
NGC5887         &  18.58 &  11.54 &  10.71 &   9.12 &   0.47 &   2.85 &   3.52 &   1.55 &   1.92\\
UGC10264        &  12.99 &   7.25 &   8.25 &   7.33 &   1.77 &   3.43 &   3.93 &   1.97 &   1.49\\
MCG0-41-5       &  15.59 &   8.53 &   9.55 &   8.05 &   1.27 &   3.26 &   3.64 &   1.84 &   1.66\\
IC4229          &   5.97 &   3.71 &   4.83 &   5.30 &   3.06 &   3.59 &   4.54 &   2.07 &   1.15\\
UGC10005        &   8.13 &   3.49 &   5.53 &   4.69 &   3.39 &   3.98 &   4.44 &   2.38 &   1.24\\
A1301-03        &   3.64 &   2.40 &   3.30 &   3.62 &   3.84 &   3.91 &   5.22 &   2.25 &   1.04\\
UGC10306        &  16.54 &  10.89 &  10.14 &   9.25 &   0.57 &   2.80 &   3.67 &   1.55 &   1.74\\
MK502           &  13.18 &   5.91 &   8.07 &   6.93 &   1.79 &   3.29 &   3.71 &   2.04 &   1.46\\
IC1235          &   4.98 &   2.71 &   4.31 &   3.98 &   5.36 &   5.00 &   5.69 &   2.64 &   1.15\\
IC1248          &   6.79 &   3.73 &   5.02 &   5.01 &   2.85 &   3.48 &   4.39 &   2.22 &   1.17\\
\end{longtable}
\noindent All equivalent wavelengths are in units of \AA.

\end{document}